\documentclass[aps,prb,twocolumn,showpacs,superscriptaddress]{revtex4-1}
\usepackage{amsfonts,amsmath,amssymb}
\usepackage[colorlinks=true,linkcolor=blue,pagecolor=blue,filecolor=blue,menucolor=blue,urlcolor=blue,citecolor=blue,anchorcolor=blue]{hyperref}%
\usepackage[dvips]{graphics}
\usepackage{graphicx}
\usepackage{bm}
\usepackage{color}

\begin{document}

\title{Electrically controllable spin filtering based on superconducting helical states}
\author{I. V. Bobkova}
\affiliation{Institute of Solid State Physics, Chernogolovka, Moscow
  reg., 142432 Russia}
\affiliation{Moscow Institute of Physics and Technology, Dolgoprudny, 141700 Russia}
\author{A. M. Bobkov}
\affiliation{Institute of Solid State Physics, Chernogolovka, Moscow reg., 142432 Russia}

\date{\today}

\begin{abstract}
The magnetoelectric effects in the surface states of the 3D TI are extremely strong due to the full spin-momentum locking. Here the microscopic theory of S/3D TI bilayer structures in terms of quasiclassical Green's functions is developed. On the basis of the developed formalism it is shown that the DOS in the S/TI bilayer manifests giant magnetoelectric behavior and, as a result, S/3D TI heterostructures can work as non-magnetic fully electrically controllable spin filters. It is shown that due to the full spin-momentum locking the amplitudes of the odd-frequency singlet and triplet components of the consensate wave function are equal. The same is valid for the even frequency singlet and triplet components. We unveil the connection between the odd-frequency pairing in S/3D TI heterostructures and magnetoelectric effects in the DOS.
\end{abstract}

\maketitle

\section{Introduction}

Spintronics is a research field, which combines magnetism and electronics \cite{zutic2004,awschalom2007,eschrig15,linder15}. Classically it is based on the opportunity of ferromagnetic materials to provide spin-polarized
currents \cite{johnson93,jedema01,urech06,huang07,jonker07,giazotto03}. However, by now it is realized
that the non-magnetic materials with strong coupling between the electron momentum and its spin give a possibility to build
purely electrically controlled elements for spintronics applications. It is predicted that an unpolarized bias current
should create a net spin polarization due to
spin-momentum coupling for both the topologically protected TI surface states \cite{burkov10,culcer10,yazyev10,bobkova16} and
the spin-orbit coupled materials \cite{aronov89,edelstein90,kato04,silov04,edelstein95,edelstein05,malshukov08,bergeret16,bobkova17}. This is called by the direct magnetoelectric effect and takes place as in normal, so as in superconducting systems, where is it dissipationless.

Manipulation by the spin degrees of freedom is best when this spin polarization and spin splitting
of the electron energy bands are very strong. If one thinks about globally magnetic system, the strongest spin differentiation is detected
in half‐metals\cite{muller08,wan05,pickett98,nie08,kobayashi98,son06,ueda98}. But if we are interested in non-magnetic materials, which are able to produce the strongest magnetoelectic effects, the candidates can be found among the surfaces of semiconductors\cite{gierz09,riley14,nitta97}, metals\cite{lashell96,koroteev04,hoesch04}, metalalloys\cite{ast07},
topological insulators and semimetals\cite{hasan10,wan11}, also the bulk non-magnetic half-metals have been reported in the literature \cite{liu17}.

Here we focus on the surface states of the 3D topological insulator (TI). The most important for us property of this system is the spin-momentum locking:
the spin of the TI Dirac surface state lies in-plane, and is locked at
right angles to the carrier momentum, see Fig.~\ref{system}(a). Therefore, the surface states of 3D TI can be viewed as an example of the non-magnetic half-metal or a helical metal.
Examples of TI materials include $\rm {Bi_{1–x}Sb_x}$ \cite{hsieh08}, $\rm {Bi_2Se_3}$, $\rm {Bi_2Te_3}$ and $\rm {Sb_2Te_3}$ \cite{zhang09,hsieh09,chen09}. Spin-momentum locking has been probed
by spin-resolved photoemission\cite{hsieh09,souma11,pan11} and polarized optical
spectroscopic techniques\cite{mciver12}.

An electric current in TI creates a net spin polarization, the amplitude
and orientation of which are controlled by the current.
This property has been predicted by theory\cite{burkov10,culcer10,yazyev10} and measured in transport experiments \cite{li14,li16}.

Recently, proximity-induced superconductivity in
surface states of a TI has been realized and demonstrated in electronic transport
measurements of several heterostructures involving a TI and
superconductor\cite{williams12,sacepe11,veldhorst12,qu12,zareapour12,xu14}.

Here we demonstrate that due to the spin-momentum locking nature of the quasiparticle spectrum the proximity-induced superconducting state in TI can be a source of highly spin-polarized electric currents. The superconductors in the Zeeman-split regime have already been proposed
as sources of highly spin polarized currents \cite{huertas-hernando02,giazotto08}. For our system the great advantage is that the value and the direction of the current polarization can be fully controlled electrically by the applied supercurrent without need to apply an external magnetic field.

The paper is organized as follows. In Sec.~\ref{simp_model} on the basis of the simplified model we discuss how the applied supercurrent leads to the spin-split DOS in the S/TI heterostructure and in Sec.~\ref{spin_filter} it is proposed how this property can be exploited to create the supercurrent-controllable spin filter. Sec.~\ref{micr} is devoted to a more strict consideration in the framework of a microscopic model, where the superconductivity in the TI surface states is induced by the proximity to the conventional superconductor. In particular, the necessary theoretical framework is derived in subsection \ref{micr1}, the DOS and the supercurrent-controllable spin filtering effect are discussed in subsection \ref{micr2} and subsection \ref{micr3} is devoted to the discussion of the spin structure and symmetry of the proximity-induced condensate wave function in the supercurrent driven regime and its connection to the DOS. Our conclusions are presented in Sec.~\ref{conclusions}.

\section{Spin-resolved DOS in supercurrent-driven S/3D TI heterostructure}
\label{simp_model}

\begin{figure}[!tbh]
  \centerline{\includegraphics[clip=true,width=3.2in]{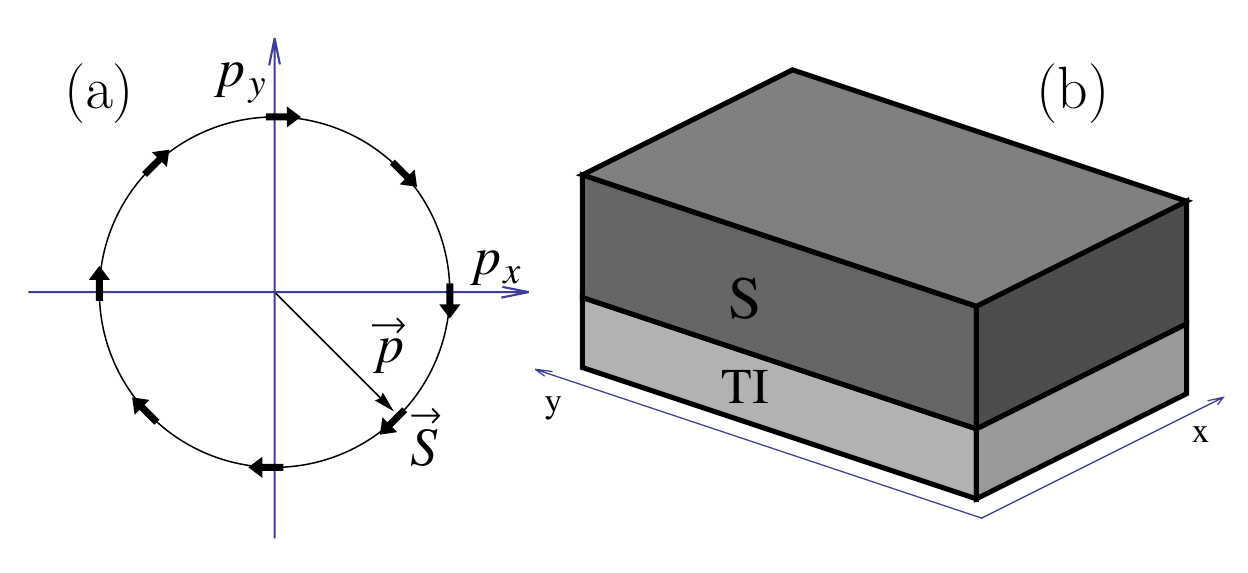}}
   \caption{(a) Fermi-surface of the TI surface states. The quasiparticle spin $\bm S$ is locked at the right angle to its momentum $\bm p$. (b) Sketch of the system under consideration. The thick top superconductor S induces the superconductivity in the surface layer of the 3D topological insulator TI.}
\label{system}
\end{figure}

We consider a planar interface between a thin singlet s-wave
superconductor (S) film and the TI surface, as depicted in Fig.~\ref{system}(b). Such a hybrid system was first considered by Fu
and Kane\cite{fu03}. At first we follow their approach, where the proximity effect on
the TI surface is described by a phenomenological singlet order parameter. The corresponding hamiltonian of the TI surface states takes the form
\begin{eqnarray}
H=\int d^2 r \Biggl\{\Psi^\dagger (\bm r)\bigl[-iv_F(\bm \nabla_{\bm r}\times \hat z)\bm \sigma - \mu +V_{imp}(\bm r)- \nonumber \\
\bm h \bm \sigma \bigr]\Psi(\bm r)+\Delta(\bm r)\Psi^\dagger_\uparrow (\bm r) \Psi^\dagger_\downarrow (\bm r) + \Delta^*(\bm r)\Psi_\downarrow (\bm r) \Psi_\uparrow (\bm r) \Biggr\},~~~~
\label{ham}
\end{eqnarray}
where $\Psi^\dagger(\bm r)=(\Psi^\dagger_\uparrow(\bm r),\Psi^\dagger_\downarrow(\bm r))$ is the creation operator of an electron at the superconducting TI surface, $\hat z$ is the unit vector normal to the surface of TI, $\bm \sigma = (\sigma_x, \sigma_y, \sigma_z)$ is a vector of Pauli matrices in spin space and $\bm h = (h_x, h_y, 0)$ is an in-plane exchange field. We allow for nonmagnetic impurity scattering potential $V_{imp}=\sum \limits_{\bm r_i}V_i \delta(\bm r - \bm r_i)$, which is of a Gaussian form $\langle V(\bm r)V(\bm r')\rangle = (1/\pi \nu \tau)\delta(\bm r - \bm r')$ with $\nu=\mu/(2\pi v_F^2)$.

We consider a situation in which the chemical potential $\mu$ is the largest energy scale in the system and hence a quasiclassical approximation is the well-suited framework to describe the system. The transport Eilenberger-type equations for the quasiclassical Green's function have been derived in Refs.~\onlinecite{zyuzin16} and \onlinecite{bobkova16}. We are only interested in the retarded component, because we investigate only DOS and the related physical quantities in the equilibrium situation. The full retarded quasiclassical Green's function $\check g^R(\bm n_F, \bm r, \varepsilon)$ is a $4 \times 4$ matrix in the direct product of the spin and particle-hole spaces, but its spin structure is determined by the projection onto the conduction band of the TI surface states and, therefore, one can write:
\begin{equation}
\check g^R(\bm n_F, \bm r, \varepsilon)=\hat g^R(\bm n_F, \bm r, \varepsilon)\frac{(1+\bm n_\perp \bm \sigma)}{2},
\label{spin_structure}
\end{equation}
where $\bm n_F=\bm p_F/p_F=(n_{F,x},n_{F,y},0)$ is a unit vector directed along the quasiparticle trajectory and $\bm n_\perp=(n_{F,y},-n_{F,x},0)$ is a unit vector perpendicular to the quasiparticle trajectory and directed along the quasiparticle spin, which is locked to the quasiparticle momentum for the helical conducting band of the TI surface states. Then the quasiclassical Green's function $\hat g^R(\bm n_F, \bm r, \varepsilon)$ is a $2 \times 2$ in the particle-hole space and obeys the following equation:
\begin{eqnarray}
-iv_F \bm n_F \bm \nabla \hat g^R = \Bigl[ \varepsilon \tau_z - \hat \Delta + \bm h \bm n_\perp \tau_z + \frac{i\langle \hat g \rangle}{2 \tau}, \hat g \Bigr],
\label{eilenberger}
\end{eqnarray}
where $\tau_{x,y,z}$ are Pauli matrices in the particle-hole space and $\hat \Delta$ is a matrix in the particle-hole space with the following explicit structure
\begin{eqnarray}
\hat \Delta =
\left(
\begin{array}{cc}
0 & \Delta e^{i\chi} \\
-\Delta e^{-i\chi} & 0
\end{array}
\right)
\label{Delta}
\end{eqnarray}
with $\chi$ standing for the phase of the superconducting order parameter.

We analyze and compare the ballistic and diffusive cases here. The Green's function in the ballistic limit can be found from Eq.~(\ref{eilenberger}) at $\tau \to \infty$, while in the diffusive limit $\varepsilon, \Delta, h \ll \tau^{-1}$ the quasiparticle motion is randomized by the impurity scattering and the Green's function can be expanded through the first two harmonics:
\begin{equation}
\label{GF_harmonics}
\hat g^R(\bm n_F)=\hat g^R_0  + \bm n_F \hat {\bm g}^R_a,
\end{equation}
where the zeroth (isotropic) harmonic obeys the Usadel equation\cite{zyuzin16,bobkova16}:
\begin{eqnarray}
i D \hat {\bm \nabla}(\hat g^R_0 \hat {\bm \nabla} \hat g^R_0) = \Bigl[ \varepsilon \tau_z - \hat \Delta, \hat g^R_0 \Bigr],
\label{usadel}
\end{eqnarray}
where $D$ is the diffusion coefficient and $\hat {\bm \nabla}\hat X=\bm \nabla \hat X - (i/v_F)(h_x \hat y - h_y \hat x)[\tau_z, \hat X]$.

The first harmonic term can be expressed via $\hat g^R_0$ as follows:
\begin{equation}
\label{GF_diffusive_first}
\hat {\bm g}^R_a=-2 v_F \tau \hat g^R_0 \hat {\bm \nabla} \hat g^R_0.
\end{equation}

Further our goal is to investigate the DOS and spin-resolved DOS in the S/TI bilayer system in the presence of the supercurrent flowing through the system along the arbitrary direction in the $(x,y)$-plane. For this purpose we find the retarded Green's function. In the ballistic case the solution of Eq.~(\ref{eilenberger}) takes the form:
\begin{eqnarray}
\hat g^R = \frac{-i}{\sqrt{|\Delta|^2-(\varepsilon+h_{eff})^2}}
\left(
\begin{array}{cc}
\varepsilon+h_{eff} & -\Delta e^{i\chi} \\
\Delta e^{-i\chi} & -(\varepsilon+h_{eff})
\end{array}
\right), ~~~~
\label{ballistic_GF}
\end{eqnarray}
where we introduce  $h_{eff}=\bm h \bm n_{\perp}-v_F(\bm n_F \bm \nabla \chi)/2$. Physically it is a projection of the effective exchange field onto the direction of the quasiparticle spin for a given quasiparticle trajectory (see below).

In the diffusive case it is convenient to express the Green's function in terms of the so-called $\theta$-parametrization\cite{belzig99}:
\begin{equation}
\label{GF_diffusive_theta}
\hat g^R_0=e^{i\chi \tau_z/2}\Bigl[\tau_z \cosh \theta + i \tau_y \sinh \theta \Bigr]e^{-i\chi \tau_z/2},
\end{equation}
\begin{equation}
\label{GF_diffusive1_theta}
\hat {\bm g}^R_a \bm n_F=2i \tau h_{eff}e^{i\chi \tau_z/2}\Bigl[ 2 \tau_z \sinh^2 \theta + i \tau_y \sinh 2\theta \Bigr]e^{-i\chi \tau_z/2},
\end{equation}
and $\theta$ should be found from the following equation:
\begin{equation}
\label{theta}
\varepsilon \sinh \theta + \Delta \cosh \theta +i D \Bigl(\frac{\bm \nabla \chi}{2}-\frac{1}{v_F}(h_x \hat y - h_y \hat x)\Bigr)^2 \sinh 2 \theta = 0.
\end{equation}

The DOS can be calculated via the upper left element (normal part) of
\begin{equation}\label{GF_matrix}
\hat g^R=
\left(
\begin{array}{cc}
g^R & f^R \\
\tilde f^R & -g^R
\end{array}
\right)
\end{equation}
as follows:
\begin{equation}
N(\varepsilon)={\rm Re}\langle g^R \rangle.
\label{DOS}
\end{equation}

The spin-resolved DOS corresponding to the spin along the direction $\hat {\bm l}$ takes the form:
\begin{equation}
N_{\hat {\bm l}}(\varepsilon)=\frac{1}{2}{\rm Re}\Bigl[\langle g^R \rangle + l_x \langle g^R n_{F,y} \rangle - l_y \langle g^R n_{F,x} \rangle\Bigr].
\label{spin_DOS}
\end{equation}

It is worth noting here that this quantity does not represent a true density of states for a given spin, because the spin direction is different for different momentum directions and, moreover, is not conserved in the presence of impurities. Nevertheless, it is this quantity that enters the transport properties of the system as the usual spin-resolved DOS does. Therefore it can be viewed as an effective spin-resolved DOS and it makes sense to discuss it.

Further we investigate DOS and the spin-resolved DOS, which is of major interest for us. We assume that the supercurrent $\bm j_s$ is applied to the system. This supercurrent results in the corresponding phase gradient $\bm \nabla \chi$ of the effective order parameter, which is the only essential parameter for our consideration. Although the supercurrent flows mainly in the top superconductor, the resulting phase gradient is the same as for the top superconductor, so as for the effective order parameter in the TI surface layer, as it is shown in Sec.~\ref{micr1}. We assume no applied exchange field in the system. However, as it can be seen from Eqs.~(\ref{ballistic_GF}), (\ref{GF_diffusive1_theta}) and (\ref{theta}), in case of the helical metal, realized in the surface states of the TI, the effect of the supercurrent on the Green's function and, therefore, on the DOS is fully equivalent to the effect of the exchange field, applied in the perpendicular to the current direction. They enter all the equations only via the effective exchange field $\bm H_{eff}=(h_x-v_F(\partial_y \chi)/2, h_y+v_F(\partial_x \chi)/2),0)$: as $\hat z \times \bm H_{eff}=h_x \hat y - h_y \hat x - v_F \bm \nabla \chi / 2 $ in Eq.~(\ref{theta}) or as $h_{eff} = \bm H_{eff} \bm n_\perp$. Consequently, all the effects, discussed below, can be caused as by the supercurrent, so as by the exchange field. We focus on the supercurrent here because it allows for the fully electrical control of the effects without need to apply a magnetic field to the system. The effective exchange field $\bm H_{eff}=(-v_F(\partial_y \chi)/2, v_F(\partial_x \chi)/2,0)$, produced by the supercurrent, is perpendicular to it. It can reach the value, comparable to the value of the gap of the top superconductor for the currents close to the depairing current. The equivalence between the exchange field and the supercurrent is violated beyond the simple phenomenological model, discussed here, but for the low-energy physics it is practically not important. We address this issue in Sec.~\ref{micr}.

\begin{figure}[!tbh]
  \centerline{\includegraphics[clip=true,width=2.8in]{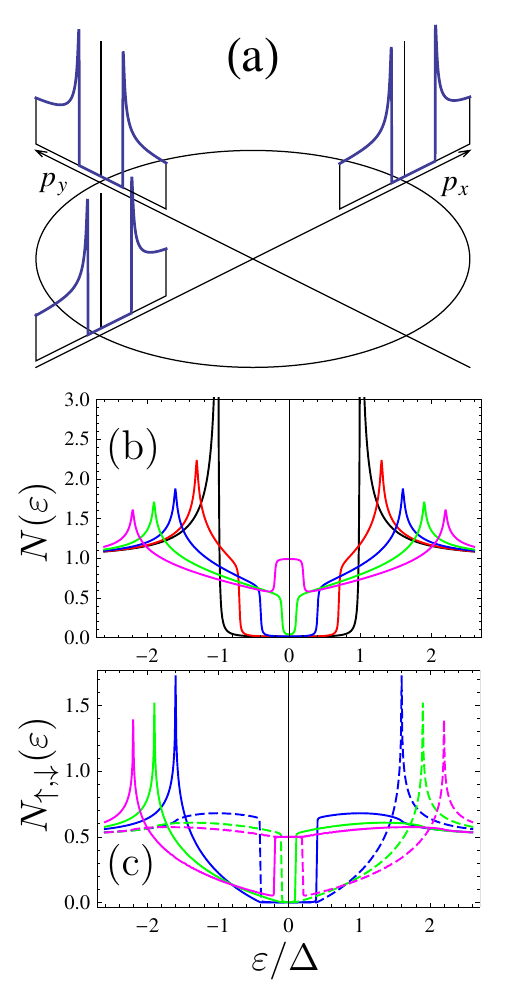}}
   \caption{DOS for the ballistic system. (a) Momentum-resolved DOS. The circle means the Fermi surface and the energy axis is directed out of this Fermi surface and normally to it at each momentum direction. (b) Momentum-averaged DOS as a function of the quasiparticle energy. $H_{eff}/\Delta=0$ (black line); $0.3$ (red); $0.6$ (blue); $0.9$ (green); $1.2$ (pink). (c) Spin-resolved DOS as a function of the quasiparticle energy. $H_{eff}/\Delta=0.6$ (blue); $0.9$ (green); $1.2$ (pink). The quantization axis is along the $y$-axis, while $\bm j_s=j_s \hat x$. The spin-up DOS $N_\uparrow$ is plotted by solid lines, and the spin-down DOS $N_\downarrow$ - by dashed lines.}
\label{DOS_ballistic}
\end{figure}

The DOS, calculated for the ballistic system according to Eqs.~(\ref{DOS}) and (\ref{ballistic_GF}), is presented in Fig.~\ref{DOS_ballistic}(b). But at first it is instructive to look at the momentum-resolved DOS, shown in Fig.~\ref{DOS_ballistic}(a). It is seen that for a given momentum direction the DOS is just shifted with respect to zero energy, analogously to the case of a definite spin subband in conventional Zeeman-split superconductors. The value of the shift is determined by the momentum direction and is maximal for the momenta parallel to the supercurrent, while the DOS is symmetric for the momenta perpendicular to it. The momentum averaging of the DOS leads to the results presented in Fig.~\ref{DOS_ballistic}(b). The gradual shrinking and closing the gap upon the increase of $H_{eff}$ is similar to the case of the Zeeman-split superconductor. We would like to note that in bulk superconductors the superconductivity at zero temperature is suppressed by the exchange field higher than the Clogston-Chandrasekhar limit of superconductivity $h>\Delta/\sqrt 2$. But here the superconductivity is induced by proximity to the bulk superconductor with the large gap $\Delta_0$. Then the proximity induced order parameter $\Delta$ depends of the S/TI interface transparency (see Sec.~\ref{micr}) and typically is considerably smaller than $\Delta_0$. In this situation it makes sense to investigate the case $H_{eff}>\Delta$, as it is done in our work and is presented in Fig.~\ref{DOS_ballistic}.

The spin-resolved DOS for the ballistic system is plotted in Fig.~\ref{DOS_ballistic}(c). The quantization axis is chosen along the $y$-axis and is directed perpendicular to the supercurrent (which flows along the the $x$-axis) The $y$-axis is just the direction of the effective exchange field in this case and the spin polarization of the DOS is maximal along this direction. We see that analogously to the case of a conventional Zeeman-split superconductor\cite{fulde73} there is an energy window, where the difference $N_\uparrow-N_\downarrow$ is very high. The reversal of the supercurrent direction leads to the interchange $N_\uparrow \leftrightarrow N_\downarrow$.

This polarization is generated by the nonmagnetic supercurrent. The effect is not possible in conventional superconductors and S/N bilayers. It only takes place due to the spin-momentum locking property of the TI surface state. This magneto-electric behavior of the DOS should also be observed for any spin-orbit coupled superconductors and superconducting hybrid systems, but it is much weaker there, because two different helical subbands are present in such materials and partially compensate the effects of each other.

\begin{figure}[!tbh]
  \centerline{\includegraphics[clip=true,width=2.8in]{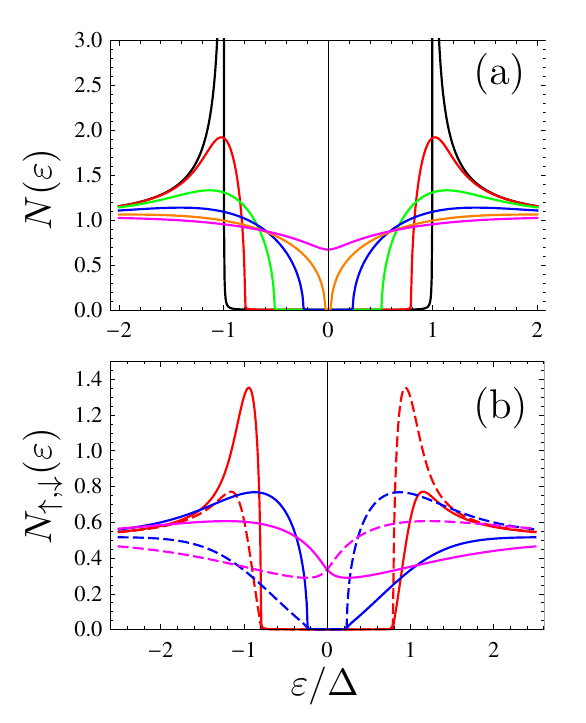}}
   \caption{DOS for the diffusive system. $\Delta \tau = 0.15$. (a) DOS as a function of the quasiparticle energy. $H_{eff}/\Delta=0$ (black line); $0.6$ (red); $1.2$ (green); $1.8$ (blue); $2.4$ (tan); $3.0$ (pink). (b) Spin-resolved DOS as a function of the quasiparticle energy. $H_{eff}/\Delta=0.6$ (red); $1.8$ (blue); $3.0$ (pink). The spin-up DOS $N_\uparrow$ is plotted by solid lines, and the spin-down DOS $N_\downarrow$ - by dashed lines.}
\label{DOS_diffusive}
\end{figure}

In order to investigate the influence of the impurities on this magneto-electric behavior of the DOS, we have calculated the DOS and the spin-resolved DOS in the diffusive case. It appears that the current-induced spin polarization of the DOS is robust against impurities. The DOS and spin-resolved DOS for the diffusive case, calculated according to Eqs.~(\ref{DOS}), (\ref{spin_DOS}), (\ref{GF_harmonics}), (\ref{GF_diffusive_theta}), (\ref{GF_diffusive1_theta}) and (\ref{theta}), are presented in Fig.~\ref{DOS_diffusive}. It is seen that qualitatively the results for the spin-resolved DOS are the same as for the ballistic case and the energy windows of very high DOS polarization survive. It is worth to note here that in the ballistic case the gap in the energy spectrum is closed at $H_{eff}=\Delta$, as it is well-known for conventional Zeeman-split superconductors. At the same time for the diffusive case the gap can still be opened at $H_{eff}>\Delta$ because, as it follows from Eq.~\ref{theta}, the parameter, which controls the smearing of the gap in this case, is $\sim D H_{eff}^2/v_F^2 \sim \tau H_{eff}^2$. This parameter can be smaller than $\Delta$ even at $H_{eff}/\Delta >1$ provided that the impurity scattering is strong $\Delta \tau \ll 1$.

The results discussed here are closely connected to the supercurrent-induced electron spin polarization, predicted \cite{bobkova16} for the homogeneous superconduting TI surface states and S/TI/S Josephson junctions. Here we demonstrate that the origin of this spin polarization is the rebuilding of the quasiparticle DOS under the applied supercurrent.

\section{N/3D TI/S heterostructure as an electrically controllable spin filter}
\label{spin_filter}

In this section we demonstrate that the discussed above magneto-electric behavior of the DOS can be exploited for the creation of fully electrically controllable spin filters on the basis of S/TI bilayers.

\begin{figure}[!tbh]
  \centerline{\includegraphics[clip=true,width=2.5in]{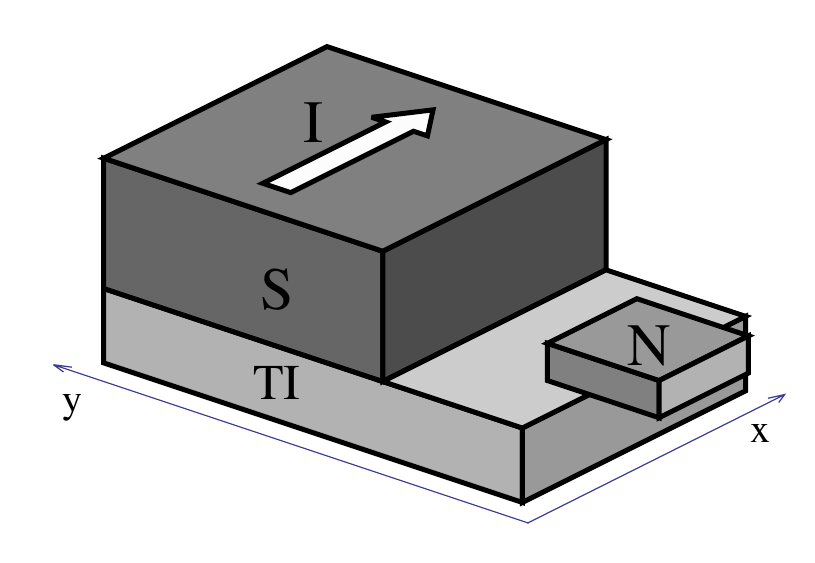}}
   \caption{Sketch of the possible setup for the realization of the electrically controllable spin filter. The spin polarized current flows though the TI/N interface in response to the voltage bias $V$ applied between the normal electrode N and the S/TI bilayer.}
\label{filter_setup}
\end{figure}

The system under investigation is shown in Fig.~\ref{filter_setup}. The additional normal electrode is attached via the tunnel junction to the surface states of the TI in the close vicinity of the top superconductor, forming a N/TI junction. The distance between the edge of the superconductor and the electrode should not exceed the superconducting coherence length in order to have the proximity-induced superconductivity in the contact region. To simplify calculations we assume that the DOS in the TI surface states in the region of the N/TI junction has just the same form as in the homogeneous S/TI bilayer, considered above. Surely, this assumption is not quite correct, because there is a decay of the proximity-induced superconductivity upon going away from the S/TI interface. But at the distances smaller than the superconducting coherence length the qualitative behavior of the Green's function survives.

We assume that the transparency of the N/TI junction is very low in order to suppress the Andreev reflection processes at the N/TI interface. These processes are detrimental for the considered spin-filtering effect because they transfer electric current via the interface in the form of opposite-spin Cooper pairs. In this tunnel limit we calculate the electric current via  the N/TI interface $I_\sigma$, corresponding to the spin $\sigma = \pm 1$ along the chosen quantization axis ($y$-axis), in the framework of the standard tunneling hamiltonian approach:
\begin{eqnarray}
I_\sigma = \frac{G}{e} \int \limits_{-\infty}^{\infty} d \varepsilon N_\sigma (\varepsilon)(f_N(\varepsilon-eV)-f_{TI}(\varepsilon)),
\label{I_spin}
\end{eqnarray}
where $G$ is the conductance through the junction in the normal state, and $f_{N/TI}(\varepsilon)$ are the Fermi
distribution functions of electrons inside the normal electrode and the TI surface states, respectively.

\begin{figure}[!tbh]
  \centerline{\includegraphics[clip=true,width=3.6in]{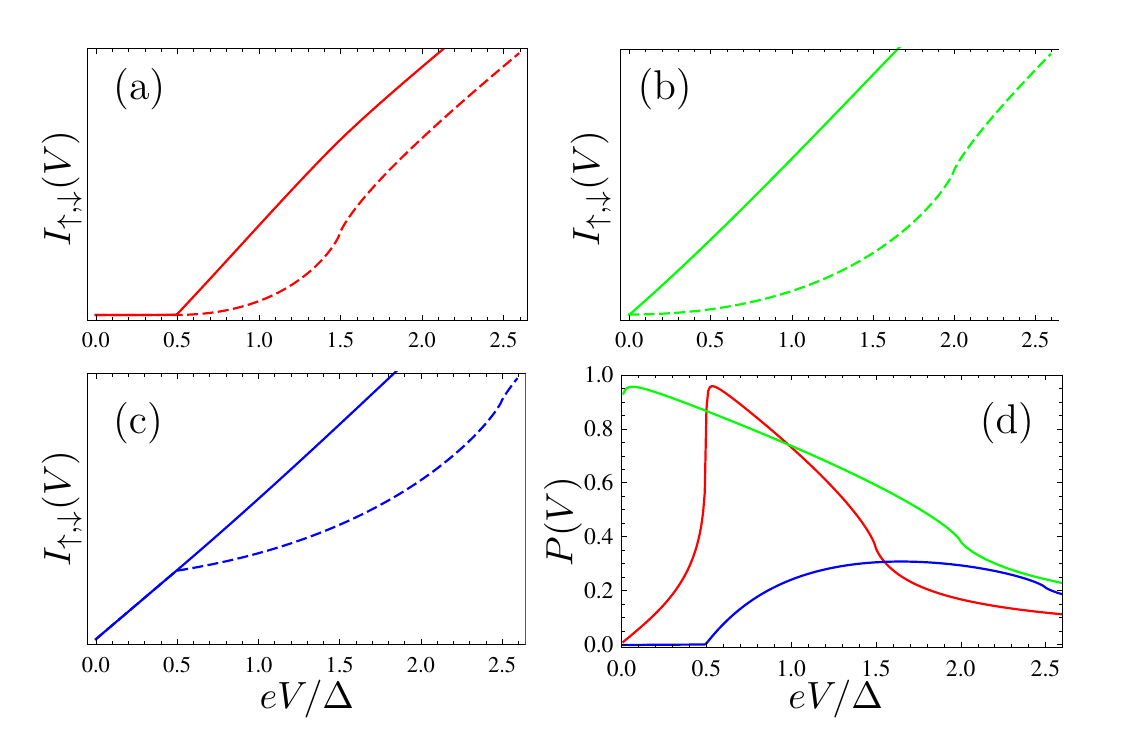}}
   \caption{Ballistic system. (a)-(c) $I_{\uparrow}$ (solid lines) and $I_{\downarrow}$ (dashed lines) as functions of $eV/\Delta$ for $H_{eff}/\Delta=0.5$ (a); $1.0$ (b) and $1.5$ (c). (d) Degree of the current polarization for all $H_{eff}$, considered in panels (a)-(c). Different colors correspond to the same $H_{eff}$, as in panels (a)-(c).}
\label{current_ballistic}
\end{figure}

The electric currents $I_{\uparrow,\downarrow}$ as functions of the voltage $V$, applied to the N/TI junction, are shown in Figs.~\ref{current_ballistic}(a)-(c) for the ballistic system. Fig.~\ref{current_ballistic}(d) demonstrates the degree of the spin polarization of this current $P=(j_\uparrow - j_\downarrow)/(j_\uparrow + j_\downarrow)$. It is seen that the degree of the current polarization is rather high for the voltage corresponding to the energy window of highly-polarized DOS. When the value of the supercurrent and, therefore, the effective exchange field is increased, the corresponding energy window shrinks. As a result, the degree of the current polarization declines for larger supercurrents.

\begin{figure}[!tbh]
  \centerline{\includegraphics[clip=true,width=3.6in]{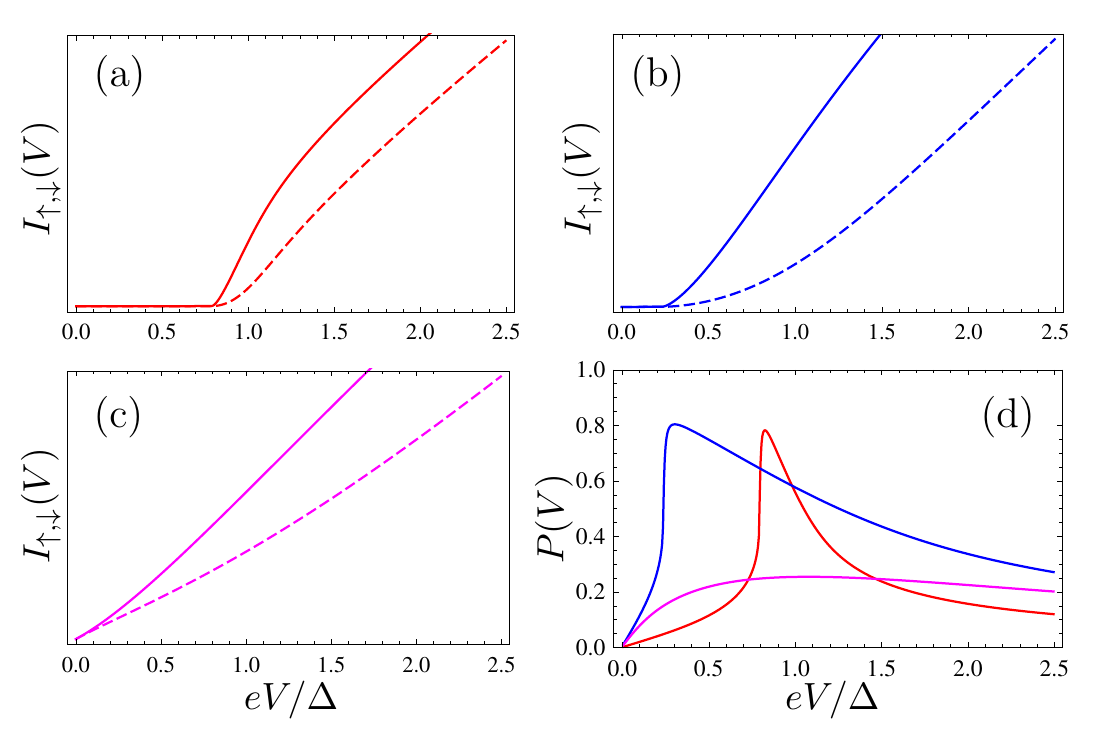}}
   \caption{Diffusive system. $\Delta \tau = 0.15$. (a)-(c) $I_{\uparrow}$ (solid lines) and $I_{\downarrow}$ (dashed lines) as functions of $eV/\Delta$ for $H_{eff}/\Delta=0.6$ (a); $1.8$ (b) and $3.0$ (c). (d) Degree of the current polarization for all $H_{eff}$, considered in panels (a)-(c).}
\label{current_diffusive}
\end{figure}

\begin{figure}[!tbh]
  \centerline{\includegraphics[clip=true,width=2.8in]{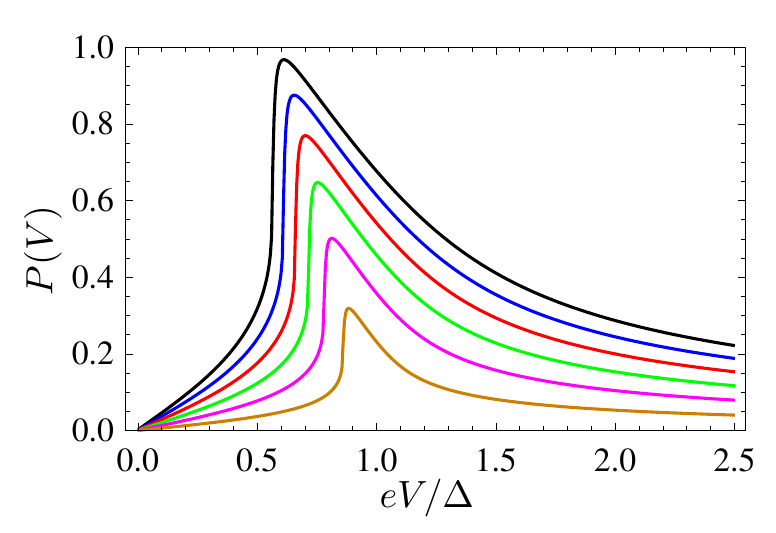}}
   \caption{Degree of the current polarization as a function of  $eV/\Delta$.  Different curves correspond to different values of impurity strength $\Delta \tau = 0.03$ (tan line), $0.06$ (pink), $0.09$ (green), $0.12$ (red), $0.15$ (blue), $0.18$ (black).  $H_{eff}=1.0 \Delta$.}
\label{current_impurity}
\end{figure}

The results for the diffusive case are qualitatively the same and are presented in Fig.~\ref{current_diffusive}. It is seen that high degree of the current  polarization can be only achieved in case if the gap in the quasiparticle spectrun is not smeared. This is valid as for the ballistic, so as for the diffusive case. For the diffusive case it is also interesting to investigate the degree of the current polarization as a function of the impurity strength at a fixed value of $H_{eff}$. The results are presented in Fig.~\ref{current_impurity}. It is seen that the degree of the current polarization declines as the impurity strength grows. This is because the spin polarization of the DOS in the diffusive case is only due to odd in momentum part of the Green's function $g_a^R \bm n_F$, and this odd in momentum part vanishes as $\tau \to 0 $ according to Eq.~(\ref{GF_diffusive_first}). In other words, the strong impurity scattering makes the Green's function to become more and more isotropic, and the nonzero spin dependence of the DOS is impossible for the fully isotropic Green's function under the condition of full spin-momentum locking in the quasiparticle spectrum.

In the opposite limit of larger $\Delta \tau$ the Usadel approximation fails. The intermediate limit $\Delta \tau \sim 1$ is more difficult for calculations, and in the clean limit $\Delta \tau \gg 1$ the results were discussed above. We do not expect any qualitative differences in spin filtering properties for the intermediate impurity strength regime $\Delta \tau \sim 1$ because the current polarizations, calculated as for the clean, so as for the diffusive cases are very similar.

Since the spin-up and spin-down DOS interchange if the direction of the supercurrent is reversed, one can reverse the direction of the current polarization electrically. The setup discussed above does not allow for obtaining the highly spin-polarized electric current with an arbitrary direction of the polarization. But this can be achieved on the basis of a little bit more complicated system. From the engineering point of view it is possible to make a setup, which allows for the supercurrent in the vicinity of the TI/N interface to flow in an arbitrary required direction. Then it is possible to obtain an arbitrary directed (perpendicular to the supercurrent) polarization of the TI/N tunnel current.

The reservoir for the spins, which are transferred from the TI to the N electrode, is the bulk top superconductor. We assume that the spin relaxation rate in the TI is faster then the rate of the tunneling through the N/TI interface. In this case the spin relaxation processes are able to maintain the equilibrium distribution of the quasiparticles in the TI surface states and our approach is applicable. If the spin relaxation is slower than the tunneling rate, the distribution function in the TI becomes nonequilibrium and spin-dependent. In this case an additional consideration of the kinetic equation is required to find the distribution function and describe the spin-filtering effect correctly.

\section{Proximity induced superconductivity in the supercurrent driven regime}
\label{micr}

All the consideration of the previous sections is based  on the phenomenological model of the proximity-induced superconductivity, where it is assumed that in the surface states of the TI there is a singlet order parameter $\Delta$. On the other hand, microscopic
approaches allow for a more general description of the proximity effect in terms of the
Green functions of the superconductor. The corresponding microscopic approaches in terms of the Gor'kov Green's functions technique were developed as for the TI surface states tunnel coupled to the superconductor \cite{stanescu10,potter11,tkachov13}, so as for various low-dimensional systems \cite{potter11,lutchyn12,golubov04,tkachov04,tkachov05,fagas05,kopnin11,stanescu11,rahimi17}. Here we derived quasiclassical Eilenberger type equations for the surface states of the TI, where the proximity effect is taken into account in terms of the
Green functions of the superconductor. This approach allows for calculation of the real-space quasiclassical Green's function in the surface states of the TI as for a homogeneous S/TI bilayer, described above phenomenologically, so as for an inhomogeneous system, which includes parts  with proximity-induced superconductivity.

\subsection{Theoretical framework}
\label{micr1}

We treat the TI surface states as a 2D system, while the top superconductor is assumed to be conventional 3D material. Assuming tunneling
coupling between the systems, we can write the Hamiltonian of such a bilayer as follows
\begin{equation}
H=H_{I}+H_S + H_T,
\label{ham_gen}
\end{equation}
where
\begin{eqnarray}
H_I = \int d^2 r \Biggl\{\Psi^\dagger (\bm r)\bigl[-iv_F(\bm \nabla_{\bm r}\times \hat z)\bm \sigma - \mu  \nonumber \\
+V_{imp}(\bm r)-\bm h \bm \sigma \bigr]\Psi(\bm r)\Biggr\}~~~~~~
\label{HI}
\end{eqnarray}
is the hamiltonian of the TI surface state,
\begin{eqnarray}
H_S = \int d^2 r dz \Biggl\{\hat c^\dagger \bigl[-\frac{\bm \nabla_{\bm r}^2+\partial_z^2}{2m_S}+V_{imp}(\bm r) - \mu_S -\bm h \bm \sigma \bigr] \hat c  \nonumber \\
+\Delta(\bm r,z)c^\dagger_\uparrow  c^\dagger_\downarrow  + \Delta^*(\bm r,z) c_\downarrow  c_\uparrow \Biggr\}~~~~~~
\label{HS}
\end{eqnarray}
is the hamiltonian of the top superconductor and
\begin{eqnarray}
H_T = \int d^2 r d^2 r' \Biggl\{\hat c^\dagger(\bm r,z=0) \hat t(\bm r, \bm r') \Psi (\bm r')+  \nonumber \\
\Psi^\dagger(\bm r') \hat t(\bm r', \bm r) \hat c (\bm r, z=0)\Biggr\}~~~~~~
\label{HT}
\end{eqnarray}
describes
the tunneling coupling between the two systems. Here $\Psi^\dagger(\bm r)=(\Psi^\dagger_\uparrow(\bm r),\Psi^\dagger_\downarrow(\bm r))$ is the creation operator of an electron at the TI surface, as before. $\hat c^\dagger (\bm r,z) = (c^\dagger_\uparrow(\bm r,z),c^\dagger_\downarrow(\bm r,z))$ is the creation operator of an electron in the superconductor. $\bm r$ is a 2D vector in plane of the TI surface and $z$ is the out-of-plane coordinate. $\hat t(\bm r, \bm r')$ is a hopping element between the TI and the superconductor. It is a $2 \times 2$ matrix in spin space and its particular form is discussed below.

Introducing the matrix Gor'kov Green's function $\check G^R(\bm r, \bm r', \varepsilon)$ for the TI surface states, as it has been done in Ref.~\onlinecite{bobkova16}, and following the derivation scheme, presented in Ref.~\onlinecite{kopnin11}, we obtain the following equation for $\check G^R$:
\begin{eqnarray}
\Bigl[ \varepsilon \tau_z + i v_F (\bm \nabla_{\bm r}\times \hat z)\bm \sigma + \mu +\bm h \bm \sigma \tau_z - \check \Sigma_{imp} - \check \Sigma_T \Bigr]\check G^R= \nonumber \\
\delta (\bm r - \bm r')\check 1, ~~~~~~~~~~~~
\label{Gor'kov_micr}
\end{eqnarray}
where $\check \Sigma_{imp}$ is the impurity self-energy, which in the Born approximation can be written as $\check \Sigma_{imp}(\bm r)=(1/\pi \tau \nu)\check G^R(\bm r, \bm r)$. $\check \Sigma_T$ is the tunneling self-energy, which comes from the coupling term between the TI and the superconductor and
\begin{eqnarray}
\check \Sigma_T \check G^R = \int d^2 r_1 d^2 r_2 d^2 r_3 \check t^\dagger (\bm r, \bm r_1)  \times ~~~~~~~ \nonumber \\
\check G^R_{S(0)}(\bm r_1, \bm r_2,z_1=z_2=0) \check t(\bm r_2, \bm r_3) \check G^R(\bm r_3, \bm r').~~~~~~~~~~~~~~
\label{tun_self_energy}
\end{eqnarray}
Here $\check t$ is the following matrix in the particle-hole space, composed of the tunneling matrices for electron and hole components:
\begin{eqnarray}
\check t(\bm r, \bm r') =
\left(
\begin{array}{cc}
\hat t(\bm r, \bm r') & 0 \\
0 & \sigma_y \hat t^*(\bm r', \bm r)\sigma_y
\end{array}
\right).
\label{t_ph}
\end{eqnarray}
The Green's function $\check G^R_{S(0)}$ is the Green's function of the top superconductor calculated for zero coupling $\hat t=0$, but with the exact self-energy terms. It should be found from the equation:
\begin{eqnarray}
\Bigl[ \varepsilon \tau_z + \frac{\bm \nabla_{\bm r}^2+\partial_z^2}{2m_S} + \mu +\bm h \bm \sigma \tau_z - \check \Sigma_{imp,S}-\Delta e^{i\chi}\tau_+ + \nonumber \\
\Delta e^{-i\chi}\tau_-\Bigr]\check G^R_{S(0)}=\delta (\bm r - \bm r')\delta(z-z')\check 1. ~~~~~~
\label{GS0}
\end{eqnarray}

Eqs.~(\ref{Gor'kov_micr}) and (\ref{GS0}) are exact, but, in principle, they are not enough for the full solution of the problem. Eq.~(\ref{GS0}) contains self-energy terms $\Delta$ and $\check \Sigma_{imp,S}$, which should be found via the full Green's function $\check G^R_{S}$ of the superconductor. This Green's function should be calculated from the equation like Eq.~(\ref{GS0}), but including the tunneling self-energies. Here we neglect the inverse proximity effect, that is the influence of the TI on the superconductor. In this case the self-energy terms $\Delta$ and $\check \Sigma_{imp,S}$ are equal to their values in the bulk superconductor and Eqs.~(\ref{Gor'kov_micr}) and (\ref{GS0}) represent the full set of equations to solve the considered problem.

Further we consider the superconductor in the presence of the applied electric supercurrent, otherwise it is homogeneous. In this case the solution of Eq.~(\ref{GS0}) takes the form:
\begin{eqnarray}
\check G^R_{S(0)}(\bm r, \bm r', z=z'=0)=\int \frac{d p_z}{2\pi}\frac{d^2 p}{(2 \pi)^2}e^{i\bm p(\bm r -\bm r')} \times \nonumber \\
e^{i \chi(\bm r)\tau_z/2}\check G^R_{S(0)}(\bm p, p_z)e^{-i \chi(\bm r')\tau_z/2},~~~~~~
\label{GS0_solution}
\end{eqnarray}
where
\begin{eqnarray}
\check G^R_{S(0)}(\bm p, p_z)=e^{-i\frac{\pi}{4}\bm M \bm \sigma}\bigl( \hat G_+ \frac{1+\sigma_z}{2}+\hat G_- \frac{1-\sigma_z}{2}  \bigr)e^{i\frac{\pi}{4}\bm M \bm \sigma}, \nonumber \\
\hat G_{\pm}=\frac{(\varepsilon -\frac{(\bm \nabla \chi) \bm p}{2m_S} \pm h)\tau_z - \xi  - \Delta i \tau_y }{(\varepsilon \pm h)^2 - \xi^2 - \Delta^2} ,~~~~~~
\label{GS0_momentum}
\end{eqnarray}
and $\xi = (\bm p^2+p_z^2)/2m_S-\mu$, $\bm M = (-h_y, h_x, 0)/h$.

Further we assume that the S/TI interface is planar and the momentum component, parallel to the interface is conserved in the hopping event. In this case the hopping element $\hat t(\bm r, \bm r')=\hat t(\bm r-\bm r')$ only depends on $\bm r -\bm r'$. Then  the tunneling self-energy term Eq.~(\ref{tun_self_energy}) can be simplified as follows:
\begin{eqnarray}
\check \Sigma_T \check G^R = \int \frac{d^2 p}{(2\pi)^2} e^{i \bm p(\bm r - \bm r')}
\check t^\dagger (\bm p) \check G^R_{S(0)} (\bm p, \bm R) \times \nonumber \\
\check t(\bm p) \check G^R(\bm p, \bm R),~~~~~~~~~~
\label{tun_self_energy1}
\end{eqnarray}
where we have turned to the mixed representation $(\bm r - \bm r'), [(\bm r + \bm r')/2 \equiv \bm R] \to \bm p, \bm R$ and neglect the terms of the order $\bm \nabla_{\bm R} \check G(\bm p, \bm R)/p_F, \bm \nabla_{\bm R} \check G^R_{S(0)}(\bm p, \bm R)/p_F$ and higher orders. To the same accuracy
\begin{eqnarray}
\check G^R_{S(0)}(\bm p, \bm R) = e^{\frac{i\chi (\bm R) \tau_z}{2}}\times \nonumber \\
\int \frac{d p_z}{2\pi}
\check G^R_{S(0)}(\bm p, p_z) e^{-\frac{i\chi (\bm R) \tau_z}{2}},
\label{GS0_quasiclassical}
\end{eqnarray}
where $\check G^R_{S(0)}(\bm p, p_z)$ is expressed by Eq.~(\ref{GS0_momentum}). Performing integrating over $p_z$ we obtain:
\begin{eqnarray}
\check G^R_{S(0)}(\bm p, \bm R) = -\frac{i}{2|v_{F,z}(\bm p)|}e^{\frac{i\chi (\bm R) \tau_z}{2}}e^{-i\frac{\pi}{4}\bm M \bm \sigma}\times \nonumber \\
\bigl( \hat g_+ \frac{1+\sigma_z}{2}+\hat g_- \frac{1-\sigma_z}{2}  \bigr)e^{i\frac{\pi}{4}\bm M \bm \sigma}e^{-\frac{i\chi (\bm R) \tau_z}{2}},
\label{GS0_quasiclassical}
\end{eqnarray}
and
\begin{eqnarray}
\hat g_{\pm}=\frac{-i\Bigl[ (\varepsilon - \frac{(\bm \nabla \chi)\bm p}{2m_S}\pm h)\tau_z - \Delta i \tau_y \Bigr]}{\sqrt{\Delta^2-(\varepsilon - \frac{(\bm \nabla \chi)\bm p}{2m_S}\pm h)^2}}.
\label{g}
\end{eqnarray}
is the standard quasiclassical matrix Green's function of the bulk superconductor. Here $v_{F,z}(\bm p)=p_{F,z}/m_S$, where $(p_{F,z}^2+p^2)/(2m_S)=\mu$.

Further we turn to the mixed representation in Eq.~(\ref{Gor'kov_micr}), subtract the analogous equation, where the Green's function $\check G^R(\bm p, \bm R)$ is on left and perform the $\xi$-integration. Following this standard procedure of the quasiclassical derivation \cite{zyuzin16}, we obtain the following Eilenberger-type equation for the quasiclassical Green's function $\check g^R(\bm n_F, \bm R, \varepsilon)$:
\begin{eqnarray}
-i\frac{v_F}{2}\left\{ \hat \eta, \bm \nabla \check g^R  \right\}=\Biggl[ (\varepsilon + \bm h \bm \sigma) \tau_z - \nonumber \\
\mu \bm n_\perp \bm \sigma -\frac{\langle \check g^R \rangle}{i\tau}-\check \Sigma_{T,q}, \check g^R \Biggr],
\label{eilenberger_micr}
\end{eqnarray}
where $\hat \eta = (-\sigma_y, \sigma_x, 0)$.
\begin{eqnarray}
\check \Sigma_{T,q} = \check t^\dagger (\bm p_F) \check G^R_{S(0)} (\bm p_F, \bm R) \check t(\bm p_F),
\label{Sigma_q}
\end{eqnarray}
and $\bm p_F$ is the Fermi momentum at the TI surface, as before.

In order to proceed further we choose the simplest and natural model for the spin structure of the hopping matrix $\hat t(\bm p)$:
\begin{eqnarray}
\hat t(\bm p)=t\frac{1+\bm n_\perp \bm \sigma}{2}.
\label{t_spin}
\end{eqnarray}
This means that only the quasiparticles with the spin directions corresponding to the helical conducting band of the TI surface state [see Fig.~\ref{system}(a)] can tunnel through the S/TI interface. In addition we choose $t$ to be real.

The spin structure of $\check g^R$ is determined by the projection onto the conduction band of the TI, Eq.~(\ref{spin_structure}). Taking this into account we obtain from Eq.~(\ref{eilenberger_micr}) the following equation for the spinless quasiclassical Green's function $\hat g^R$:
\begin{eqnarray}
-iv_F \bm n_F \bm \nabla \hat g^R = \Bigl[ (\tilde \varepsilon + \tilde{\bm h} \bm n_\perp)\tau_z - \hat {\tilde \Delta} +  \frac{i\langle \hat g \rangle}{2 \tau}, \hat g \Bigr]
\label{eilenberger_micr_spinless}
\end{eqnarray}
This equation has the same form as Eq.~(\ref{eilenberger}), obtained in the framework of the simple phenomenological model, but the quasiparticle energy term and the exchange energy term are renormalized by the diagonal in particle-hole space parts of the tunneling self-energy term:
\begin{eqnarray}
\tilde \varepsilon = \varepsilon + \frac{t^2}{4|v_{F,z}(\bm p_F)|}\Bigl[ \frac{\varepsilon-\frac{(\bm \nabla \chi)\bm p}{2m_S}+ h}{\sqrt{\Delta^2-(\varepsilon-\frac{(\bm \nabla \chi)\bm p}{2m_S}+ h)^2}}+ \nonumber \\
\frac{\varepsilon-\frac{(\bm \nabla \chi)\bm p}{2m_S}- h}{\sqrt{\Delta^2-(\varepsilon-\frac{(\bm \nabla \chi)\bm p}{2m_S}- h)^2}} \Bigr],~~~~~~~~~
\label{energy_renorm}
\end{eqnarray}
\begin{eqnarray}
\tilde {\bm h} = \bm h + \frac{t^2}{4|v_{F,z}(\bm p_F)|}\Bigl[ \frac{\varepsilon-\frac{(\bm \nabla \chi)\bm p}{2m_S}+ h}{\sqrt{\Delta^2-(\varepsilon-\frac{(\bm \nabla \chi)\bm p}{2m_S}+ h)^2}}- \nonumber \\
\frac{\varepsilon-\frac{(\bm \nabla \chi)\bm p}{2m_S}- h}{\sqrt{\Delta^2-(\varepsilon-\frac{(\bm \nabla \chi)\bm p}{2m_S}- h)^2}} \Bigr]\frac{\bm h}{h}.~~~~~~~~~~
\label{h_renorm}
\end{eqnarray}

The phenomenological order parameter in Eq.~(\ref{eilenberger}) is also substituted by the effective order parameter $\hat {\tilde \Delta}$, which originates from the Cooper pair wave function of the top superconductor and takes the form:
\begin{eqnarray}
\hat {\tilde \Delta} =
\left(
\begin{array}{cc}
0 & \tilde \Delta e^{i\chi} \\
-\tilde \Delta e^{-i\chi} & 0
\end{array}
\right)
\label{Delta}
\end{eqnarray}
\begin{widetext}
\begin{eqnarray}
\tilde \Delta = \frac{t^2}{4|v_{F,z}(\bm p_F)|}\Biggl[\Bigl(\frac{\Delta}{\sqrt{\Delta^2-(\varepsilon-\frac{(\bm \nabla \chi)\bm p}{2m_S}+ h)^2}}+
\frac{\Delta}{\sqrt{\Delta^2-(\varepsilon-\frac{(\bm \nabla \chi)\bm p}{2m_S}- h)^2}}\Bigr)+ \nonumber \\
\Biggl[\Bigl(\frac{\Delta}{\sqrt{\Delta^2-(\varepsilon-\frac{(\bm \nabla \chi)\bm p}{2m_S}+ h)^2}}-
\frac{\Delta}{\sqrt{\Delta^2-(\varepsilon-\frac{(\bm \nabla \chi)\bm p}{2m_S}- h)^2}}\Bigr)\frac{\bm h \bm n_\perp}{h}\Biggr].~~~~~~~~~~
\label{Delta_renorm}
\end{eqnarray}
\end{widetext}

It is seen that in the framework of the considered microscopic model the full equivalence between the Zeeman term and the supercurrent is lost. This is because it is absent in the top superconductor, which is not a helical metal. The Zeeman term simply shifts the energy there, while the supercurrent acts as a momentum-dependent Doppler shift and smears all the DOS features. However, at low energies $\varepsilon, H_{eff} \ll \Delta$ to the zeroth order in $(\varepsilon,H_{eff})/\Delta$, our microscopic model reduces to the phenomenological one, discussed above, with $\tilde \varepsilon \to \varepsilon$, $\tilde h \to h$ and the proximity-induced order parameter $\tilde \Delta_{\varepsilon \to 0} \approx t^2/2|v_{F,z}|$. In this low-energy region the equivalence between the Zeeman term and the supercurrent is preserved.

\subsection{Spin-resolved DOS and spin-filtering effect}
\label{micr2}

\begin{figure}[!tbh]
  \centerline{\includegraphics[clip=true,width=3.0in]{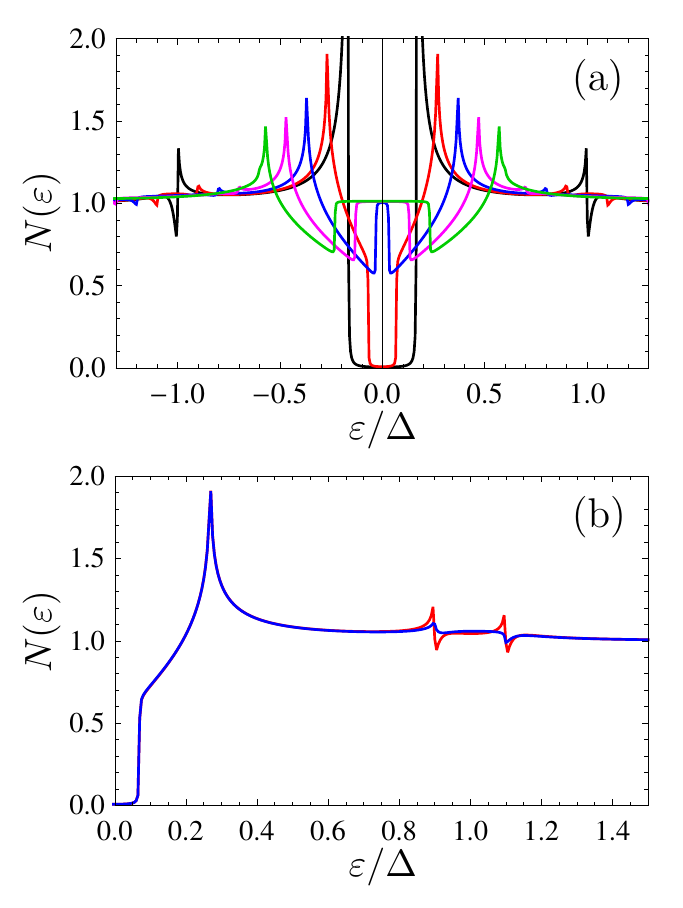}}
   \caption{(a) DOS calculated in the framework of the microscopic model. Different curves correspond to different values of the supercurrent with $H_{eff}=0$ (black), $0.1$ (red), $0.2$ (blue), $0.3$ (pink), $0.4$ (green) in units of $\Delta$, the true exchange field $h=0$. (b) DOS at $j_S \neq 0$ (corresponding to $H_{eff}=0.1\Delta$) and $h=0$ (blue line) in comparison to DOS at $j_S=0$ and $h=0.1 \Delta$ (red line). $t^2/(2|v_{F,z}|)=0.2\Delta$ for the both panels.}
\label{DOS_micr1}
\end{figure}

We focus on the ballistic limit here. The Green's function is expressed by Eq.~(\ref{ballistic_GF}) with the substitution $\tilde \varepsilon$, $\tilde h$ and $\tilde \Delta$ for $\varepsilon$, $h$ and $\Delta$, respectively.
The DOS and spin-resolved DOS are calculated according to Eqs.~(\ref{DOS}) and (\ref{spin_DOS}), respectively. The results for the DOS are plotted in Fig.~\ref{DOS_micr1}. All the energies are normalized to the gap value of the top superconductor $\Delta$. The results, presented in Fig.~\ref{DOS_micr1} correspond to the case of low-transparency S/TI interface $t^2/2|v_{F,z}|=0.2 \Delta$. Panel (a) represents the DOS when the true Zeeman field is zero and only a supercurrent is applied to the system. Different curves are for different values of the supercurrent. It is seen that the DOS behavior in the low-energy region is practically the same as in the framework of the phenomenological model. The differences between the phenomenological and the microscopic models are only in the high energy region $\varepsilon \sim \Delta$, where the additional peaks, arising from the gap of the top superconductor, appear. The spin-resolved DOS manifests qualitatively the same behavior and we do not show the corresponding figures.

Panel (b) of Fig.~\ref{DOS_micr1} demonstrates the DOS for nonzero supercurrent, corresponding to $H_{eff}=0.1 \Delta$ (blue curve), in comparison to the DOS for the true Zeeman field $h=0.1 \Delta$ and no applied supercurrent (red curve). It is seen that there is a full equivalence between the supercurrent-driven and field-driven results in the low-energy region. The difference can be seen only for energies $\varepsilon \sim \Delta$, where the peaks, induced by the top superconductor gap are smeared by the supercurrent as compared to the case of the applied Zeeman field.

\begin{figure}[!tbh]
  \centerline{\includegraphics[clip=true,width=3.6in]{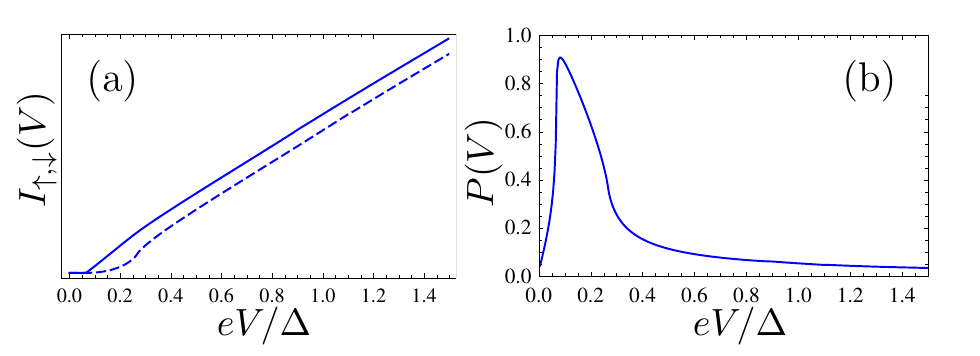}}
   \caption{Spin-filtering properties, calculated in the framework of the microscopic model. $t^2/(2|v_{F,z}|)=0.2\Delta$. Supercurrent-driven regime. (a) $I_{\uparrow}$ (solid line) and $I_{\downarrow}$ (dashed line) as functions of $V/\Delta$ for $H_{eff}=0.1 \Delta$. (b) Degree of the current polarization for the same $H_{eff}$.}
\label{current_micr1}
\end{figure}

Further we investigate the spin-filtering properties of the system in the framework of this microscopic model. The results are summarized in Fig.~\ref{current_micr1}. Comparing this Figure to Fig.~\ref{current_ballistic}, which presents the same quantities, calculated on the basis of the phenomenological model, we see that they are practically the same, the additional features at $eV \approx \Delta$ are too small to be seen in Fig.~\ref{current_micr1}.

\begin{figure}[!tbh]
  \centerline{\includegraphics[clip=true,width=3.6in]{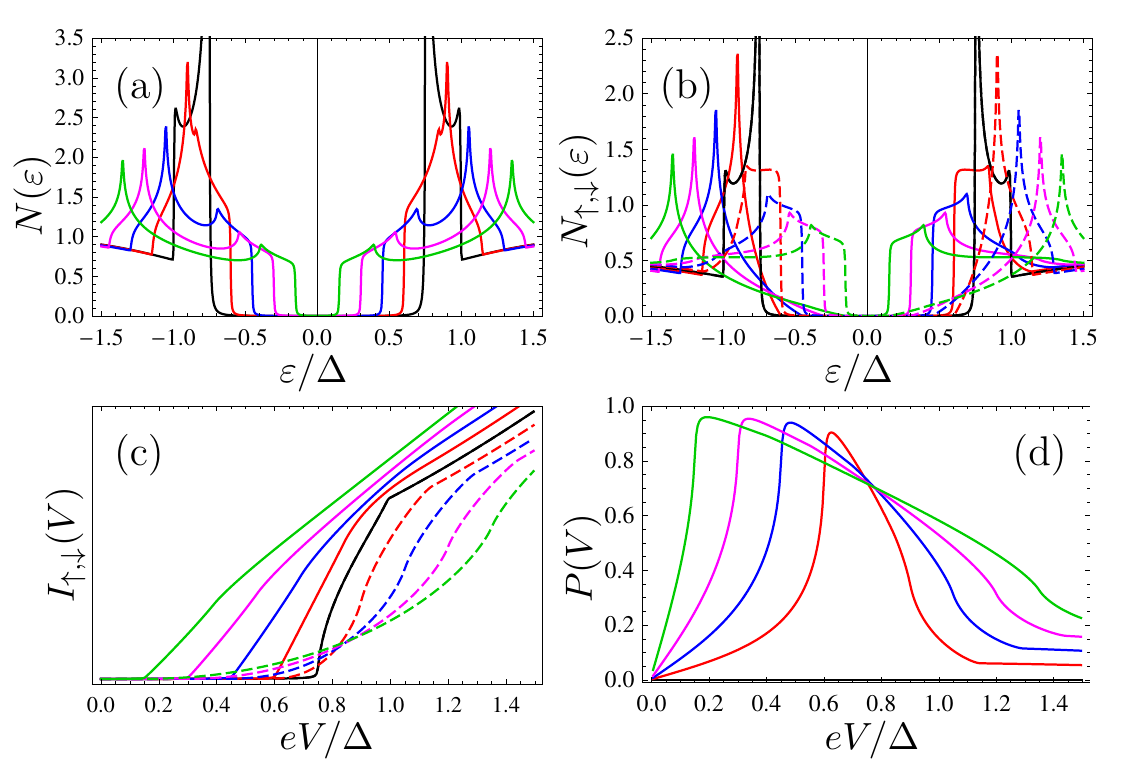}}
   \caption{Results, calculated in the framework of the microscopic model, high-transparency case $t^2/(2|v_{F,z}|)=2.0\Delta$. (a) DOS for $H_{eff}=0$ (black line), $0.15$ (red), $0.3$ (blue), $0.45$ (pink), $0.6$ (green). (b) Spin-resolved DOS for the same $H_{eff}$. As before, the spin-up DOS is plotted by solid lines and the spin-down DOS - by dashed lines. (c) $I_{\uparrow}$ (solid lines) and $I_{\downarrow}$ (dashed lines) as functions of $V/\Delta$ for the same $H_{eff}$. (d) Degree of the current polarization for the same $H_{eff}$.}
\label{DOS_current_micr2}
\end{figure}

All the results, presented above, are calculated for a low-transparent S/TI interface with $t^2/2|v_{F,z}|\ll \Delta$. Further we demonstrate the behavior of the DOS, spin-resolved DOS and the spin-filtering properties for the opposite case of high-transparent S/TI interface with $t^2/2|v_{F,z}|= 2 \Delta$. It is known that for such a high-transparent case in non-helical systems the proximity-induced gap becomes very close to the gap of the top superconductor, but still slightly smaller than it \cite{kopnin11}. Here we observe the same picture. The DOS and spin-resolved DOS are shown in panels (a) and (b) of Fig.~\ref{DOS_current_micr2}, respectively. Qualitatively their behavior is close to the results obtained in the framework of the phenomenological model, but the quantitative fit by this phenomenological model is not possible.

The spin-polarized current for the high-transparent case is shown in Fig.~\ref{DOS_current_micr2}(c) for different $H_{eff}$, and the corresponding current polarization is plotted in Fig.~\ref{DOS_current_micr2}(d). Again, qualitatively the results are very close to the results of the phenomenological model, but quantitatively here there are two clearly visible characteristic features in the current-voltage characteristics, which come from the proximity-induced gap and the top superconductor gap, respectively.

\subsection{Spin structure and symmetry of the proximity-induced condensate wave function}
\label{micr3}

Cooper pair amplitudes
are classified into four types according to their behavior with respect to frequency,
momentum (parity), and spin \cite{eschrig07,tanaka07,tanaka07_2}. Using a "frequency/spin/momentum" notation: ESE (even frequency, spin singlet, even momentum),
OSO (odd frequency, spin singlet, odd momentum), ETO (even frequency, spin triplet, odd momentum)  and OTE (odd frequency, spin triplet, even momentum).

The four symmetry states above exhaust all possibilities compatible with Fermi statistics and
the Pauli exclusion principle. The usual spin singlet, s-wave BSC superconductor is of ESE type, while the spin triplet, p-wave superfluid formed in
3He \cite{leggett75,vollhardt90} or in Rashba superconductors \cite{gorkov01,alicea10} is of ETO type. OTE type was first considered by Berezinskii \cite{berezinskii74} in connection
with early research on superfluid 3He. OSO type was introduced in connection with
unconventional superconductors \cite{balatsky92,abrahams95,dahal09}. There are also suggestions for realization of odd-frequency superconductivity (OSO or OTE) in different models \cite{eschrig15}.

But up to now an odd-frequency superconductor  with global spontaneous symmetry breaking has not been found in nature.
At the same time, the odd-frequency superconductivity frequently takes place locally in some spatial regions, e.g. near interfaces, line defects,
or inclusions \cite{eschrig15,linder17}. As concerns the proximity-induced odd-frequency superconductivity at the TI surface states, the Zeeman term-induced OTE \cite{burset15} as well as the electric current-induced OTE component \cite{black-schaffer12} were predicted. The OSO component was also discussed in the presence of the perpendicular exchange field and taking into account the hexagonal warping of the Fermi surface \cite{vasenko17}. The odd-frequency superconductivity was also predicted due to the combination of spin-momentum locking and translational symmetry breaking in S/N heterostructures based on the edge channels of 2D TI \cite{cayao17}.

Here we demonstrate that in case if the chemical potential is placed far from the Dirac point and the Fermi surface consists of only one helical band, the property of spin-momentum locking dictates that the amplitudes of the odd-frequency singlet OSO and triplet OTE components are equal. The same is valid for the even frequency components ESE and ETO.

Indeed, the spin structure of the anomalous Green's function $\hat f^R(\bm n_F, \bm R, \varepsilon)$ is determined by Eq.~(\ref{spin_structure}). The momentum-dependent spinless amplitude $f^R(\bm n_F, \bm R, \varepsilon)$ of this Green's function should be found from Eq.~(\ref{eilenberger_micr_spinless}). For an arbitrary impurity strength it can be written as $f^R(\bm n_F)=f^R_s(\bm n_F)+f^R_a(\bm n_F)$, where $f^R_{s,a}(\bm n_F)=(f^R(\bm n_F) \pm f^R(-\bm n_F))/2$ are its symmetric and antisymmetric parts with respect to $\bm n_F \to -\bm n_F$. Then for the full anomalous Green's function $\hat f^R(\bm n_F, \bm R, \varepsilon)$, which is a wave function of the Cooper pair and determines its symmetry, we obtain:
\begin{eqnarray}
\hat f(\bm n_F, \bm R, \varepsilon) = \frac{f_s}{2} + \frac{f_s }{2}\bm n_\perp \bm \sigma +
\frac{f_a}{2} + \frac{f_a }{2} \bm n_\perp \bm \sigma.
\label{CP_symmetry}
\end{eqnarray}
Further, the symmetry of $f_{s,a}$ with respect to Matsubara frequency $\omega_n$ follows from the general symmetry relations for the quasiclassical Green's functions: $\hat f^{tr}(\bm n_F, \omega_n)=\sigma_y \hat f(-\bm n_F, -\omega_n) \sigma_y$ \cite{serene83}. Substituting into this relation the spin structure of $\hat f$ Eq.~(\ref{spin_structure}) and turning to momentum-symmetric and antisymmetric components $f_{s,a}$ we obtain that $f_s(\bm n_F, \omega_n)=f_s(\bm n_F, -\omega_n)$ and $f_a(\bm n_F, \omega_n)=-f_a(\bm n_F, -\omega_n)$.

\begin{figure}[!tbh]
  \centerline{\includegraphics[clip=true,width=2.8in]{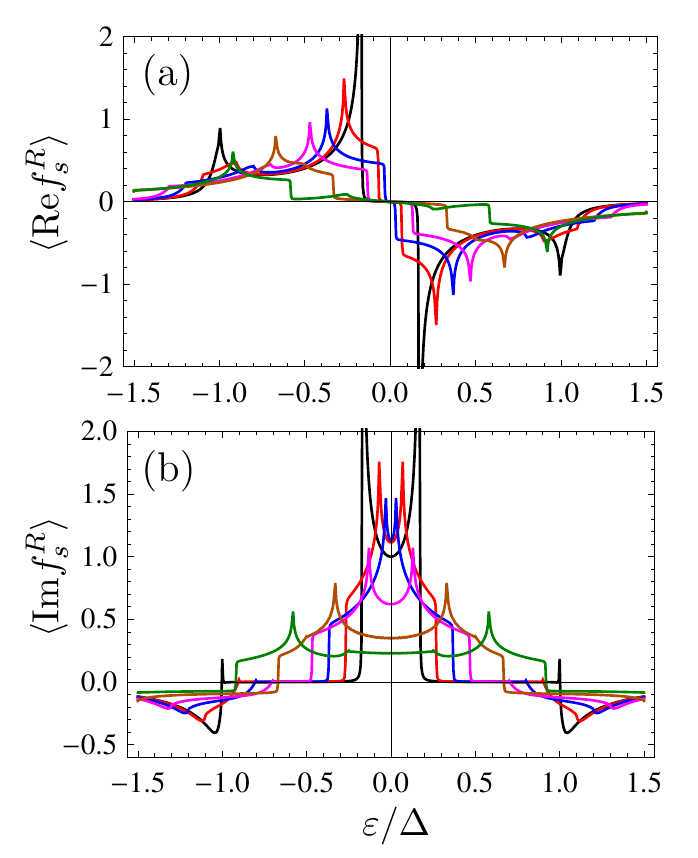}}
   \caption{(a)Averaged over momentum directions $\langle {\rm Re}f_s^R \rangle $ as a function of $\varepsilon$. (b) $\langle {\rm Im}f_s^R \rangle$. For the both panels $\varepsilon$ is normalized to the gap of the top superconductor $\Delta$. Different curves correspond to different $H_{eff}=0$ (black), $0.1\Delta$ (red), $0.2$ (green), $0.3$ (pink), $0.5$ (tan), $0.75$ (green). Microscopic model with $t^2/(2|v_{F,z}|)=0.2\Delta$.}
\label{condensate1}
\end{figure}

\begin{figure}[!tbh]
  \centerline{\includegraphics[clip=true,width=2.8in]{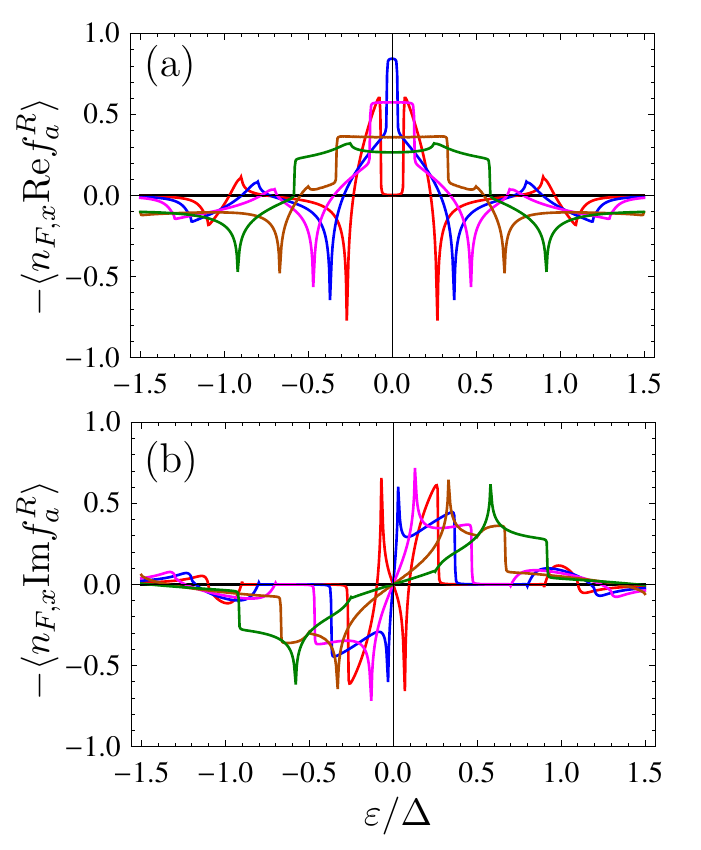}}
   \caption{(a) Averaged over momentum directions $-\langle n_{F,x}{\rm Re}f_a^R \rangle $ as a function of $\varepsilon$. (b) $-\langle n_{F,x}{\rm Im}f_a^R \rangle$. The colors marking $H_{eff}$ and the other parameters are the same as in Fig.~\ref{condensate1}.}
\label{condensate2}
\end{figure}

Therefore, we can conclude that the first two terms in Eq.~(\ref{CP_symmetry}) represent the ESE and ETO amplitudes, respectively, while the second two terms correspond to OSO and OTE amplitudes. It is seen that $f_a$, which is nonzero only in presence of $h_{eff} \neq 0$ in our system, is the measure of the both types of the odd-frequency correlations simultaneously. For the ballistic case the particular expressions for the even-frequency amplitude $f_s$ and odd-frequency amplitude $f_a$ can be obtained immediately from Eq.~(\ref{ballistic_GF}) with the substitution $\tilde \varepsilon$, $\tilde h$ and $\tilde \Delta$ for $\varepsilon$, $h$ and $\Delta$, respectively. In order to demonstrate the behavior of even-frequency and odd-frequency components panels (a) and (b) of Fig.~\ref{condensate1} represent the real and imaginary parts of $\langle f_s^R \rangle$ (averaged over momentum directions) as functions of $\varepsilon$ for different effective exchange fields, while the real and imaginary parts of $-\langle n_{f,x}f_a^R \rangle$ are demonstrated in panels (a) and (b) of Fig.~\ref{condensate2}. At $H_{eff}=0$ the antisymmetric part $f_a^R = 0$ (black line coinciding with zero level). Upon increasing $H_{eff}$ the antisymmetric part $f_a$ appears at first in the vicinity of the coherence peaks, and after that spreads over all the energies.
It is seen that ${\rm Re}f_s^R $ is antisymmetric with respect to the energy, and ${\rm Im}f_s^R$ is symmetric. For $f_a$ the situation is the opposite. It is equivalent to the fact that in Matsubara frequencies $f_s$ is the even function and $f_a$ is the odd one.

It is interesting that the proximity-induced superconductor, discussed here, allows for investigation of the superconducting correlations at the effective Zeeman fields exceeding the superconducting gap value, what is practically not possible for singlet bulk superconductors in the presence of the Zeeman field (because at zero temperature the singlet superconductivity is destroyed for $h$ exceeding the Pauli limit $\Delta/\sqrt2$ and also because at nonzero field the transition to normal state is a first-order phase transition). It is seen from Figs.~\ref{condensate1} and \ref{condensate2} that in the regimes $H_{eff}<\Delta_g$ and $H_{eff}>\Delta_g$, where  $\Delta_g$ is the proximity-induced gap, the behavior of the condensate wave functions is qualitatively different. At $H_{eff}<\Delta_g$ the distance between the coherence peaks shrinks upon increasing $H_{eff}$, while this distance gets larger upon further increase of $H_{eff}>\Delta_g$. In this regime in the low-energy region $\varepsilon \ll \Delta$ the $\langle {\rm Im}f_s^R \rangle \sim -\langle n_{F,x}{\rm Re}f_a^R \rangle $ are approximately independent on energy, while ${\rm Re}f_s^R$ and ${\rm Re}f_a^R $ are much smaller.

It is also worth to note that the spin-resolved DOS and the odd-frequency amplitude $f_a$ are closely connected. In terms of the symmetric and antisymmetric with respect to the momentum parts of the Green's function Eq.~(\ref{spin_DOS}) can be rewritten as
\begin{equation}
N_{\hat {\bm l}}(\varepsilon)=\frac{1}{2}{\rm Re}\Bigl[\langle g_s^R \rangle + l_x \langle g_a^R n_{F,y} \rangle - l_y \langle g_a^R n_{F,x} \rangle\Bigr].
\label{spin_DOS1}
\end{equation}
It is seen that the DOS can be spin-dependent only due to $g_a^R \neq 0$. From the other hand, it can be easily obtained from the normalization condition that
\begin{equation}
g_a^R = - \frac{f_a^R \tilde f_s^R + f_s^R \tilde f_a^R}{2 g_s^R},
\label{ga_fa}
\end{equation}
that is, in the system with the spin-momentum locking the difference $N_\uparrow - N_\downarrow$ and the odd-frequency amplitude of the condensate wave function are zero or nonzero simultaneously.

\section{Conclusions}
\label{conclusions}

The microscopic theory of S/3D TI bilayer structures in terms of quasiclassical Green's functions is developed. The theory is formulated for real-space quantities and allows for treating a number of transport and inhomogeneous problems.

On the basis of the developed formalism it is shown that due to the property of full spin-momentum locking the influence of the Zeeman field and the supercurrent on the low-energy properties of the proximity-induced superconductivity are the same. The DOS in the S/TI bilayer manifests giant magneto-electric behavior and, as a result, S/3DTI heterostructures can work as non-magnetic electrically controllable spin filters.

In the presence of the in-plane Zeeman term or the supercurrent all 4 symmetry-allowed types of superconducting correlations are induced by the proximity effect in 3D TI/S heterostructures and the amplitudes of the odd-frequency singlet OSO and triplet OTE components are equal. The same is valid for the even frequency components ESE and ETO.

\begin{acknowledgments}

We thank M. A. Silaev for fruitful discussions.

\end{acknowledgments}


\end{document}